\documentclass[conference,singlespace]{IEEEtran}
\usepackage[graphics]{preview}
\usepackage{cite}
\usepackage{graphicx}
\usepackage{url}
\usepackage{algorithm}
\usepackage{algorithmic} 
\usepackage{balance}
\usepackage{hyperref}
\usepackage{amsmath}
\usepackage{array}

\begin{document}

\title{A Semi-distributed Reputation-based Intrusion Detection System for Mobile Adhoc Networks}
\author{Animesh Kr Trivedi$^1$, Rajan Arora$^1$, Rishi Kapoor$^1$, Sudip Sanyal$^1$ and Sugata Sanyal$^2$\\ $^1$Indian Institute of Information Technology, Deoghat, Jhalwa, Allahabad (U.P.), India\\ \{aktrivedi\_b03,rarora\_b03,rkapoor\_b03,ssanyal\}@iiita.ac.in\\ $^2$School of Technology and Computer Science, Tata Institute of Fundamental Research, India\\ sanyal@tifr.res.in}
\date{Nov 13, 2006}
\maketitle
\begin{abstract}
A \emph{Mobile Adhoc Network} (\textsc{manet}) is a cooperative engagement of a collection of mobile nodes without any centralized access point. The underlying concept of coordination among nodes in a cooperative \textsc{manet} has induced in them a vulnerability to attacks due to issues like 
dynamically changing network topology, cooperative algorithms and lack of centralized monitoring point. 
 We propose a semi-distributed approach towards a reputation-based \emph{Intrusion Detection System} (\textsc{ids}) that combines with the Dynamic Source Routing (\textsc{dsr}) protocol for strengthening the defense of a \textsc{manet}. Our system inherits the features of reputation from human behavior, hence making the \textsc{ids} socially inspired. It has a semi-distributed architecture as the critical observations of the system are neither spread globally nor restricted locally. 
The system assigns maximum priority to self observation by nodes for updating any reputation parameters, thus avoiding the need of a trust relationship between nodes. Our system is also unique in the sense that it features the concepts of \emph{Redemption} and \emph{Fading} with a robust \emph{Path Manager} and \emph{Monitor} system. 
Simulation studies show that \textsc{dsr} fortified with our system outperforms normal \textsc{dsr} in terms of the packet delivery ratio and routing overhead even when up to half of nodes in the network behave as malicious. Various parameters introduced such as timing window size, reputation update values, congestion parameter and other thresholds have been optimized over several simulation runs.
By combining the semi-distributed architecture and other design essentials like path manager, monitor module, redemption and fading concepts, our system proves to be robust enough to counter most common attacks in \textsc{manet}s.


\end{abstract}

\section{Introduction}
The term \emph{adhoc networks} dates back to the 1970's where an adhoc network was first setup as a part of certain defense research projects. With advances in microelectronics technology and networking protocols, it has been possible to integrate mobile nodes and various other network devices into a single unit called an \emph{adhoc} node. Further, interconnection of these nodes wirelessly is termed as an \emph{adhoc network}.\\
\textsc{manet}s are different from conventional networks.
A \textsc{manet} is formed by an autonomous system of mobile nodes that are self-configuring and have no constraints, such as a fixed infrastructure or a central administration system. Nodes in \textsc{manet}s are both routers and terminals. They are dynamic in the sense that each node is free to join and leave the network in a nondeterministic way. In addition, they do not have a clearly defined physical boundary, and therefore, no specific entry or exit point. Such a network can thus be rapidly deployed and can provide the amount of flexibility and adaptability which is otherwise unattainable under adverse circumstances. Although \textsc{manet} is a very promising technology, challenges are slowing its development and deployment. Nodes in adhoc networks are in general limited in battery power, memory and \textsc{cpu} capacity. Hence the transmission ranges of these devices are also limited and nodes have to rely on neighbor nodes in the network to route the packet to its destination. They are sometimes referred to as multihop networks, where a hop is a direct link between two nodes. Adhoc networks have found applications in emergency rescues, battlefield operations, mobile conferencing, national crisis, home and community networking, disaster recovery etc.\\
\indent The flexible structure and volatile environment of \textsc{manet}s results in significant node misbehavior. Not only does it degrade the overall network performance, but, it also becomes difficult to detect intruders on grounds of mobility and vulnerability of the nodes. Thus, there is a serious need for a robust \textsc{ids} for \textsc{manet}s.\\ 
\indent Some fundamental problems of \textsc{manet}s must be kept in mind while designing any security solution. Firstly, it is often very hard to differentiate intrusions and normal operations or conditions in \textsc{manet}s because of the dynamically changing topology and volatile physical environment. Secondly, mobile nodes are autonomous units that are capable of roaming independently in an unrestricted geographical topology. This means that nodes with inadequate physical protection can be captured, compromised or hijacked. Thirdly, decision-making in adhoc networks is usually decentralized and many adhoc network algorithms rely on the cooperative participation of all nodes. Most adhoc routing protocols are also cooperative in nature and hence can be easily misguided by false routing information. Without any counter policy, the effects of misbehavior have been shown to dramatically decrease network performance. In this paper, we propose a new technique based on reputation for efficiently solving the problem of intrusion detection.\\
The next section gives a brief background about routing related issues in \textsc{manet}s, section III entails a discussion of some related efforts which is followed by the system design overview in section IV. Section V describes the protocol and the following section VI talks about its implementation details. Simulation results and optimization procedures for parameters such as window size are given in the section VII. The last section presents some concluding remarks.

\section{BackGround}
In order to understand the nature of attacks on \textsc{manet}s, we first need to look at the routing protocols for these networks. They have been classified under two main categories - Proactive and Reactive routing protocols. \emph{Proactive} protocols work with tables that are used to store routing information and updates are triggered to propagate any information about changes throughout the whole network. The obvious advantage is that routes to any destination node are always available without the overhead of generating a \emph{route request} whenever the need for a route arises. But, an extra overhead is always a major issue before deploying a proactive routing protocol, because it generally affects the overall throughput and power usage. Destination-Sequenced Distance Vector (\textsc{dsdv}) \cite{rDSDV}, Wireless Routing Protocol (\textsc{wrp}) \cite{rWRP}, Cluster Gateway Switch Routing (\textsc{cgsr}) \cite{rCGSR} are some common examples.\\
 On the other hand, \emph{Reactive} routing protocols are \emph{on-demand} i.e. a route discovery mechanism is initiated whenever there is a need for setting up a path for communication between a source and a destination node. The source node initiates route discovery by flooding the network successively with route queries. The destination node on receiving a route request (\textsc{rreq}) addressed to it, sends back a route reply (\textsc{rrep}) message as unicast to the source node either through the discovered route or by initiating another route request. Generally, on-demand routing requires less overhead than table-driven routing; but it incurs a path discovery delay whenever a new path is needed. Dynamic Source Routing protocol (\textsc{dsr})\cite{rDSR}, Adhoc On-Demand Distance Vector Routing (\textsc{aodv}) protocol \cite{rAODV}, Temporally Ordered Routing Algorithm (\textsc{tora}) \cite{rTORA}, Associativity-Based Routing \textsc{(abr)} \cite{rABR}, Signal Stability Routing (\textsc{ssr}) \cite{rSSR}, Zone-Based Hierarchical Link State Routing Protocol (\textsc{zrp}) \cite{rZRP} are a few examples. \\

Attacks are possible on reactive protocols like \textsc{dsr} due to lack of built-in security measures and the assumption of honest coordination and cooperation among nodes and with the protocol. We will outline a few attacks by nodes below, the others are discussed in detail by Sonja Buchegger et. al. \cite{WDOG}:
\begin{itemize}
\item Dropping all packets not destined to it or performing only partial dropping. Partial dropping can be restricted to specific types, such as only data packets or route control packets or packets destined to specific nodes.
\item Sending forged routing packets, an attacker can create a so-called black hole, a node where all packets are discarded or all packets are lost.
\item Modifying the nodes list in the header of a \textsc{rreq} or a \textsc{rrep} to misroute packets and adding incorrect routes in the route cache of other nodes.
\item Decreasing the hop count \textsc{dsr(ttl)} when receiving a packet, so that the packet will never be received by the destination. This attack could be detected by the previous node in route by enhanced passive acknowledgment.
\item Initiating frequent \textsc{rreq} to consume bandwidth and energy and to cause congestion.
\end{itemize} 

\section{Related Work}
Reputation-based systems are a new paradigm and are being used for enhancing security in different areas. These systems are lightweight, easy to use and are capable of facing a wide variety of attacks.
Among these mechanisms, \textsc{core} \cite{rCORE}, \textsc{confidant} \cite{rCONF} and \textsc{ocean} \cite{rOCEAN} gain a special mention.\\
\indent Reputation based systems do not rely on the conventional use of a common secret to establish confidential and secure communication between two parties. Instead, they are simply based on each other's observations. Reputation based systems are used for enhancing security in adhoc networks as they model cooperation between the nodes which is inspired from social behavior. Such systems are used to decide whom to trust and to encourage trustworthy behavior. 
Resnick and Zeckhauser \cite{rTRUST} identify three goals for reputation systems:
\begin{itemize}
\item To provide information to distinguish between a trustworthy principal and an untrustworthy principal.
\item To encourage principals to act in a trustworthy manner
\item To discourage untrustworthy principals from participating in the service the reputation mechanism is present to protect.
\end{itemize}
\emph{Watchdog} and \emph{Path-rater} \cite{WDOG} are some essential components of any typical reputation based \textsc{ids}. Watchdog performs the activity of monitoring its neighborhood and based on these observations, pathrater ranks the available path in route cache.
Misbehavior detection and reputation-based intrusion detection may be either distributed or local. Here, fully distributed means that information regarding one's reputation change is immediately propagated in the whole network. In the latter case, called local reputation based systems, nodes are fully dependent on their personal opinion about other nodes' reputation and behavior.\\
\indent Distributed \textsc{ids} protocols rely only on first-hand information with optional second-hand information. \textsc{core}  \cite{rCORE} 
proposed by P. Michiardi and \textsc{confidant}  \cite{rCONF} proposed by Buchegger and Le Boudec fall into this category. Some basic problems with this approach of global reputation systems are:
\begin{itemize}
\item Every node has to maintain O(n) reputation information where n is number of nodes in network.
\item Extra traffic generation in reputation exchange.
\item Extra computation in accepting indirect reputation information (secondhand information) esp. Bayesian Estimation.
\item Security issues in reputation exchange such as reputation data packets can be modified.
\end{itemize} 
\indent \textsc{confidant} detects misbehaving nodes by means of observation or by \textsc{Alarm} signals from neighborhood. It aggressively informs nodes in neighborhood about misbehavior of the malicious node. The weightage of \textsc{Alarm} warning signal depends upon the level of trust of receiving node about the sending node. In addition, it uses bayesian estimation for various measures and calculation of trust and reputation and thus, the \textsc{ids} becomes complex. 
\textsc{confidant} is vulnerable to false accusations if trusted nodes lie or if several liars collude \cite{FACC}.\\
\textsc{core} \cite{rCORE} proposed by P. Michiardi et. al. uses a mechanism to enforce node
cooperation in \textsc{manet}s. In this mechanism, reputation is a measure of someone's contribution to network operations. Members that have a good reputation can use available resources while members with a bad reputation cannot, because they refused to cooperate earlier and are gradually excluded from the community.
\textsc{core} defines three types of reputation:
\begin{itemize}
\item Subjective reputation is a reputation value which is locally calculated based
on direct observation.
\item Indirect reputation is second hand reputation information which is
established by other nodes.
\item Functional reputation is related to a certain function, where each function is
given a weight as to its importance. For example, data packet forwarding
may be deemed to be more important than forwarding packets with route
information, so data packet forwarding will be given greater weight in the
reputation calculations.
\end{itemize}
\textsc{Core} reputation values range from positive (+1), through null (0), to negative
(-1). \textsc{core} suffers from the problem of unwanted consequence of good reputation,
where a good node may even wish to decrease its reputation by behaving badly to
prevent its resources being over-used. The \textsc{core} mechanism assumes that every
node will use the same reputation calculations and will also assign the same
weights to the same functions. This is a potentially inappropriate assumption in
heterogeneous adhoc networks, where devices with different capabilities and roles
are likely to place different levels of importance on different functions depending
upon \textsc{cpu} usage, battery usage etc. One can take advantage of this situation and
may perform only those functions which have higher preferences in calculating
reputation. \\
\indent The second type of \textsc{ids} may be categorized as \emph{local systems}. They solely depend upon the first hand observation of their neighbors for reputation maintenance. \textsc{ocean} \cite{rOCEAN} 
by Bansal and Baker falls into this category. In these systems, nodes make routing decisions based only on direct observations of their neighbor nodes. This eliminates most of the trust manager complexity, but, doesn't fit well to a highly mobile adhoc network. In such a network, it may be difficult for the reputation upgrading process to cope up with the node mobility and it might not be appropriate to depend solely upon personal observation. Also, using secondhand information can significantly accelerate the detection and subsequent isolation of malicious nodes in \textsc{manet}s \cite{SHAND}.

\section{System Overview}
As stated earlier \cite{rRISM}, the system design is based on the reputation paradigm and possesses a \emph{semi-distributed} nature in terms of the reputation exchange mechanism.

The term semi-distributed is used in the system observation context, which is neither restricted to the observing node nor immediately propagated to the whole network as is the case in true distributed systems. 
The system design has been kept simple keeping in mind the amount of traffic already in the network and constraints such as the critical amount of battery and computational power that individual nodes possess. The system runs on every node in the network and consists of the following modules:
\subsection{Monitor}
In wireless networks, acknowledgements are often provided at no cost, either as an existing standard part of the \textsc{mac} protocol in use (such as the link-layer acknowledgement frame defined by \textsc{Ieee} 802.11) or by a ``passive acknowledgement" (in which, a node confirms receipt at another node by overhearing the transmission from sender).
The Monitor holds the responsibility of monitoring activities in the neighborhood using \emph{\textsc{pack}}s (Passive \textsc{Ack}nowledgements) which have been provided as a feature in the \textsc{dsr} protocol specifications \cite{rDSR} as promiscuous mode. 

Every node registers all the data packets sent by it to its next hop neighbor and on overhearing packets in promiscuous mode, it  matches  those against packets registered in the queue. These packets are considered as \textsc{pack}s only if both of the following two tests 
succeed:
\begin{itemize}
\item The source address, destination address, protocol,
       identification and fragment offset fields in the IP header
       of the two packets must match, and
\item If either packet contains a \textsc{dsr} source route header, both packets
       must contain one and the value in the `segments left' field in the
       \textsc{dsr} source route header of the new packet must be less than that
      in the first packet. 
\end{itemize}
 
A crucial new parameter introduced in our system is the \emph{timing window} that is a fixed time interval. After each timing window, nodes make a log of number of packets for which they have not received acknowledgment in the form of \textsc{pack} and communicate this information to the reputation system. 
In existing reputation systems, every packet is kept waiting for its \textsc{pack} for a fixed time interval. In contrast, we use the concept of timing window, which gives us the flexibility of checking timeout on fixed intervals 
rather than checking it on the basis of each individual packet's timeout. Monitor maintains a log of activity of next neighbor for each window and sends it to the reputation manager. Depending upon its cooperation, performance and current environment conditions, reputation system updates the nodes' reputation. 
With the help of Timing Window, the system also takes into consideration  \emph{congestion state} of nodes, which shall be explained in next subsection.

\subsection{Reputation System}
Reputation system module assigns and maintains reputation of different nodes as a numeric value with a lower limit of 0 and upper limit equal to the value of $MaliciousThreshold$. Reputation of any node can change by three means, as shown in Figure~\ref{fig:SysBehav}:
\begin{figure}[h]
\centering
\includegraphics[scale=0.4]{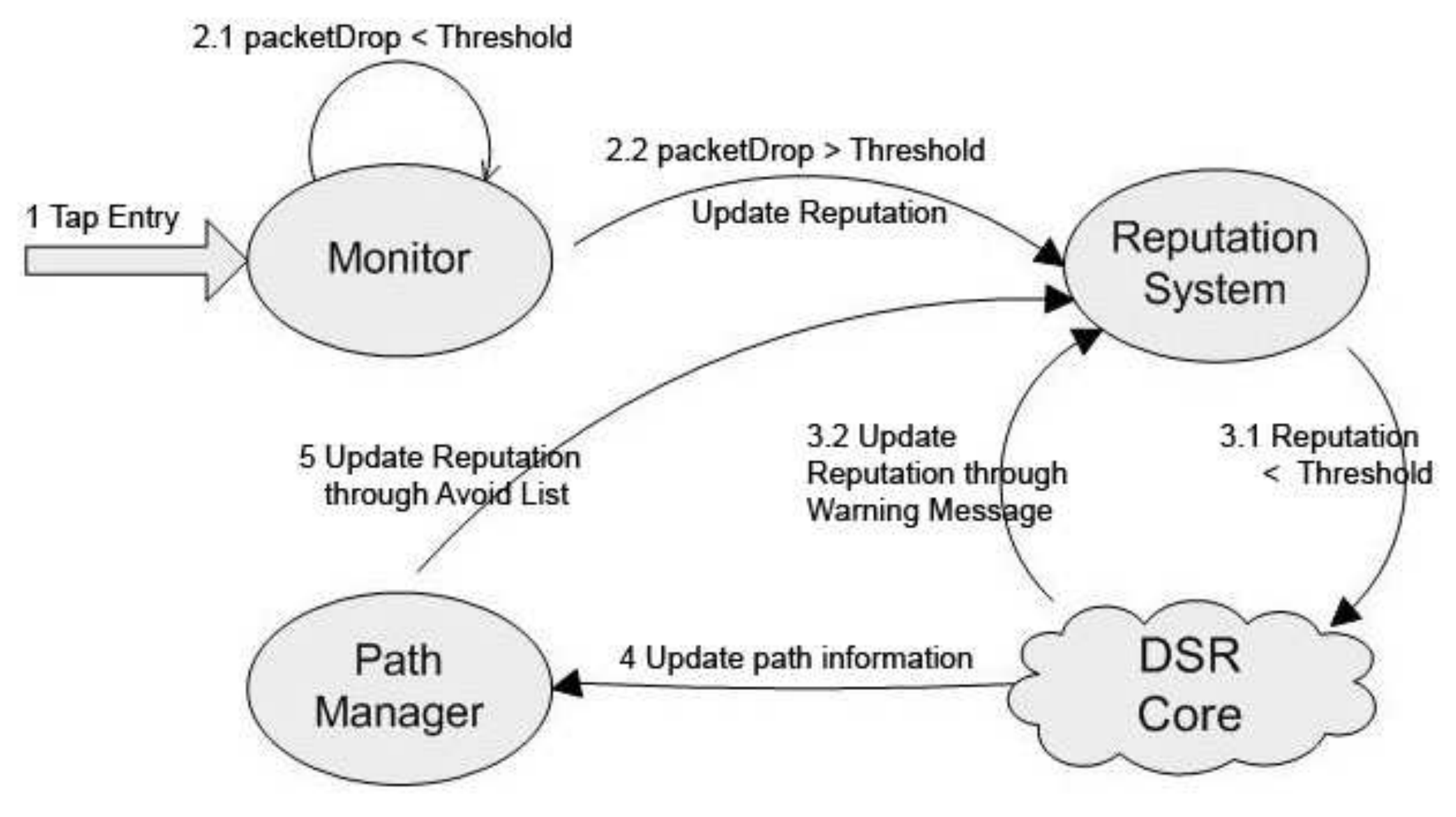}
\caption{\label{fig:SysBehav}The System Behavior}
\end{figure}
\begin{itemize}
\item  By Self observation
\item \textsc{Warning} Message, issued by neighboring nodes
\item Avoid List, appended to the \textsc{rreq/rrep} header
\end{itemize}
All three means of reputation change have some associated reputation weights with them, giving maximum weightage to self observation. The reputation is updated after every timing window and the information is communicated in a sporadic way by means of avoid lists thereby avoiding much of network overhead. 
 The concept of \emph{avoid list} is inherited from \textsc{ocean} \cite{rOCEAN}. It facilitates easy communication among nodes by putting their malicious node list in the \textsc{rreq} header. This helps in reducing the extra network traffic which would otherwise be generated while communicating this information among the peers.
A  node  may  be tagged  as  \emph{normal},  \emph{suspicious}  or  \emph{malicious}  depending  on  the reputation  value  associated  with  it. Every new node starts with a normal reputation value of zero and this reputation value may be lowered by degrading its performance or it may be incremented through the \emph{positive appraisal} feature on normal behavior. To add to the robustness and performance of our system, it is ensured that absolute value of the negative decrement awarded is larger than the positive appraisal. However at no point should the reputation of node go above zero to prevent the kind of attacks, where a node first gains positive reputation but later on depicts a malicious behavior, thereby bringing its reputation value back to the normal range. It also avoids the peculiar situation where a node may end up exhausting all its crucial resources in routing extra traffic faced due to the popularity gained by earning positive reputation. After each timing window, reputation system receives activity log of next hop neighbor from monitor with number of packets for which it has not received \textsc{pack}, which are classified as \emph{missing} or \emph{dropped} packets. The number of missing packets is then compared with the \emph{MaliciousDropThreshold} and if it is comparatively lesser, then the reputation manager gives positive performance appraisal otherwise a negative one. Unlike existing systems our system does not have a rigid MaliciousDropThreshold, we introduce the concept of \emph {congestion parameter}, which is given as:

\begin{equation}
\footnotesize{Congestion\ Parameter}\ =\ \frac{Current\ queue\ length}{Total\ queue\ length}
\end{equation}

 With the assumption that the next node is also in same congestion state as the node in contention. Misbehavior drop threshold, that is the allowed number of packet drops in a timing window is dynamically decided as:

\begin{equation}
\footnotesize{MaliciousDropThreshold} = 
\footnotesize{MaxPacketRate}\ \times \ {CongestionParameter} \times \ {WindowSize} \
\end{equation}

Whenever a new node is categorized as malicious, a warning message is spread only to its immediate neighborhood, thus protecting the network flooding with reputation update messages. This can be understood from the Figure~\ref{fig:Scenario}: 

\begin{figure}[h]
\centering
\includegraphics[scale=0.3]{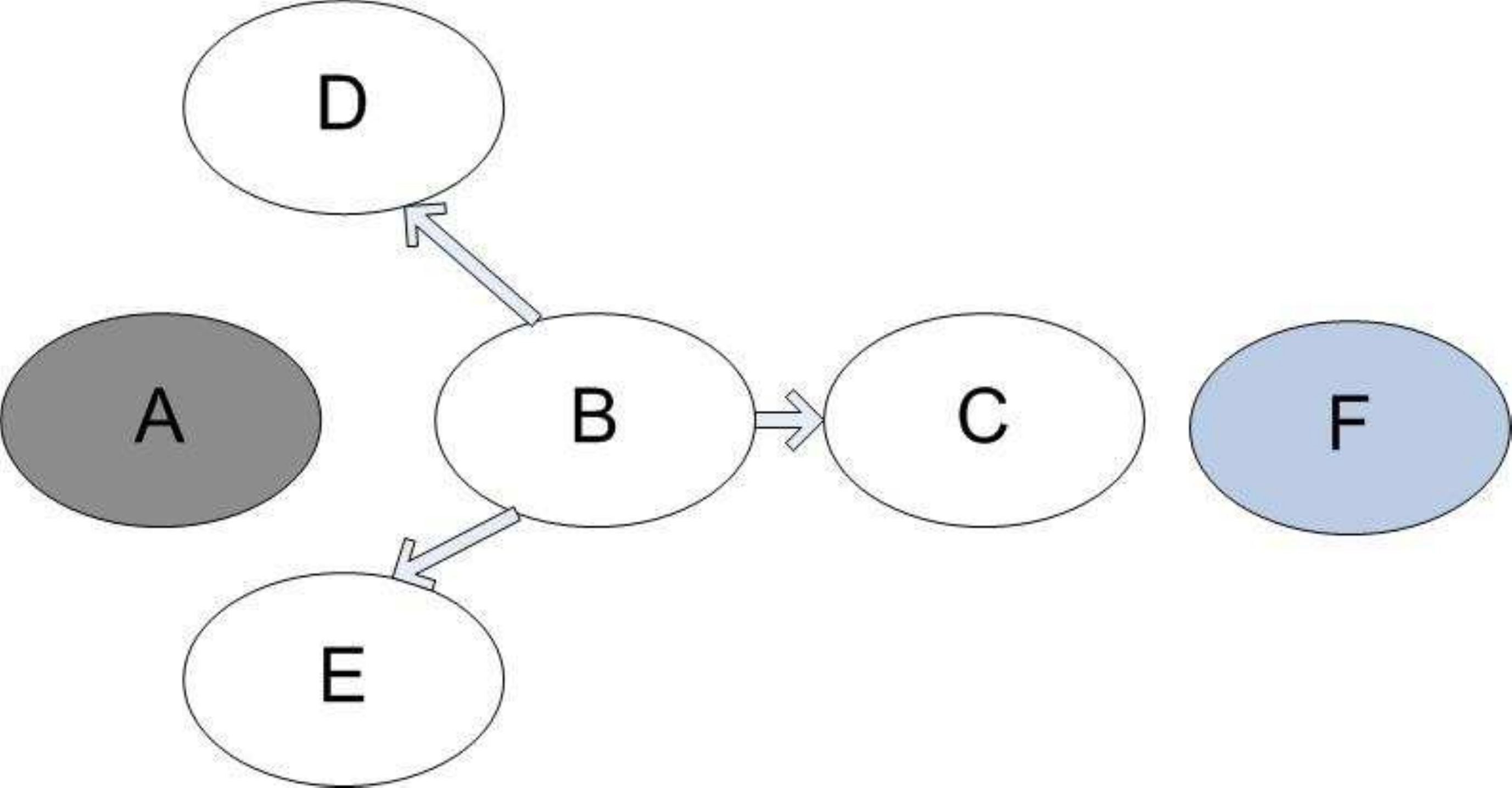}
\caption{\label{fig:Scenario}A typical network scenario}
\end{figure}

If Node B categorizes A as malicious, a Warning Message is spread to all immediate neighbor nodes: C, D and E (not to F). Nodes C, D and E on receiving a warning message decrease the reputation of node A, against which the message was originally published. Lastly, another mode of reputation updating is by means of an avoid list \cite{SHAND}.
During the route discovery phase the \textsc{rreq} sending node puts its malicious node list in \textsc{rreq} packet header and initiates discovery. When a node receives a route request packet it decreases the reputation of all those nodes quoted in the avoid list by a predefined weight. The node also appends its own malicious node list in the header and then forwards the route request packet.\\
   In order to deal with the attacks on a typical reputation system, like those of `Collusion of liars' and `false warning messages', the system has a policy that \emph{nodes can be categorized malicious only by self-observation}. It helps in nullifying attacks of the aforementioned types as the false warning messages spread by nodes can only decrease reputation of the victim nodes to a certain extent, termed as \emph{suspicious threshold}. Once this threshold is reached, the system solely depends upon self observation for making the final decision. Warning messages and avoid list are only effective above the suspicious threshold. 
Whenever a node's reputation is in the suspicious category and a deciding node receives a new warning message or an avoid list appearance for the previous one, the system performs a \emph{knock test}. Knock test is a unique test designed for checking the authenticity of a node against whom the deciding node constantly receives such information. For instance, if node A falls into the suspicious category and node B receives another warning message or an avoid list appearance corresponding to node A, the deciding node B performs a knock test over A, explained later in the protocol description section.

\subsection{Path Manager}

The path manager performs trivial path management functions in collaboration with \textsc{dsr} core. 
Path ranking is done according to path priority formula (3). Updating path-cache on various events such as those when new nodes are declared as malicious or a malicious node is taken back in network; taking decision on receiving route request or traffic from malicious nodes are a few responsibilities of the path manager.
Concept of avoid list has been added to path manager, which is a list of malicious nodes that a certain node possesses and is appended  to  the  \textsc{rreq}  header  whenever  a  route  request is  issued  by  some  node. Nodes which find themselves in avoid list do not process the packet and may simply drop it. During \textsc{rrep},  only  a  path  with  clean  nodes  is  preferred  over  those containing  suspicious  nodes and malicious nodes. Replies from such nodes are also dropped and nodes do not process request and/or forward data packets from such nodes. If during traffic flow, a new node is declared  malicious, then, all paths containing it are deleted from route cache and a route error is generated, stating that their link to the destination node has been broken. Neighbors, after receiving a route error, clear the activity log of the node which generates a route error from the current timing window. 
   The  following function  may  be used to decide the path priority if need arises: 

\begin{equation}
\footnotesize{Path\ Priority}\ \propto\ \tfrac{1}{|Min\ reputation\ of\ Node\ in\ path| \times no\ of\ hops}
\end{equation}

\subsection{Redemption And Fading}
\emph{Redemption} and \emph{Fading} are introduced in our design to allow nodes previously considered malicious to become a part of the network again. 
\textsc{manet}s run on cooperation and collaboration of peer nodes and no one gets benefited without cooperating with each other. Knock test is crucial for nodes in suspicious category and node may fail this test due to various reasons like transient link failures, congestion or resetting of the network interface etc. and once they fail this test, they are declared as malicious. To account for these problems, our system uses the \emph{fading} mechanism. After a certain inactivity period the reputation of a node is improved by a certain predefined fading rate and finally the node is moved from the malicious list to middle of suspicious category. But, the node is not given \emph{neutral rating} \cite{NTRAL} 
so that if the node again misbehaves then it is immediately put in malicious list and all transactions through that node are blocked. Here, inactivity period means no appearance in any \textsc{warning} messages or avoid list.

\section {Protocol Description}
\indent This section entails a discussion of the actual working of the system and provides the flow for various activities at different types of nodes. 
Following algorithms give a concise idea of the route discovery phase, monitoring mode and knock test feature of our system as discussed in earlier sections.

 $$SENDER$$
\begin{algorithmic} [1]
\STATE $\Rightarrow$ Generate \textsc{rreq} Packet\\
\STATE $\Rightarrow$ Pack Malicious List in \textsc{rreq} Header as Avoid List\\
\STATE $\Rightarrow$ Propagate Request
 $$OTHER\ NODES$$\\
\IF {(Own name present in Avoid List)}
\STATE $\Rightarrow$ Drop Request\\
\ELSE 
\STATE $\Rightarrow$ Scan Avoid List\\
\STATE $\Rightarrow$ Update Node's Reputation\\
\STATE $\Rightarrow$ Append its own malicious list to RREQ header avoiding repetition\\
\IF {(Node is same as Destination in \textsc{rreq})}
\STATE $\Rightarrow$ Prepare Reply\\
\ELSE 
\STATE $\Rightarrow$ Add itself in route and propagate\\
\ENDIF
\ENDIF

\end{algorithmic}

\indent The above algorithm presents a node's behavior during route establishment phase. Sender of the \textsc{rreq} just initiates the route discovery process with avoid list of malicious nodes packed in the \textsc{rreq} packet header. The remaining nodes after receiving such requests process the avoid list attached in the received \textsc{rreq} header. If a matching entry is found for their own name in the list, the node drops the request. Otherwise, the reputation of the other nodes present in the avoid list is updated. If the receiving node is the destination for which the \textsc{rreq} has been sent, then it prepares a route reply else it appends its own malicious list in the header to the existing avoid list avoiding repetitions and propagates the route request.


 $$MONITOR\ MODE$$
Self Observation-
\begin{algorithmic} [1]
\IF {(Performance is below normalThreshold)}
\STATE $\Rightarrow$ Negative reputation update\\
\ELSE 
\STATE $\Rightarrow$ Positive reputation update\\
\IF {(reputation is above 0)}
\STATE $\Rightarrow$ SET reputation = 0
\ENDIF
\ENDIF
 $$WARNING\ MESSAGE\ PROPAGATION$$
\IF {(WARNING\_MSG \&\& NEIGHBOR)}
\IF {(Reputation below Suspicious Threshold)}
\STATE $\Rightarrow$ Perform Knock Test\\
\IF {(Knock Test is Passed)}
\STATE $\Rightarrow$ Assign normal reputation\\
\ELSE 
\STATE $\Rightarrow$ declare as Malicious\\
\STATE $\Rightarrow$ spread Warning Message\\
\ENDIF
\ENDIF
\ELSE
\STATE $\Rightarrow$ decrease reputation\\
\ENDIF

\end{algorithmic} 

 The system in monitoring mode has three ways of gathering information for reputation updation:
\begin{itemize}
\item Self Observation
\item Warning Message
\item Avoid List
\end{itemize}
Some observations just monitor the neighbor with the help of \textsc{pack}. If the performance lies below the suspicious threshold, then a negative reputation update is performed over the node in consideration, otherwise a positive appraisal is given. Warning messages are only processed if they are for immediate neighbors. If the reputation of a node under consideration is below the suspicious threshold, then the knock test is performed. Otherwise, the reputation is decreased linearly. Table 2 contains actual values of these constant parameters used during system simulation.


 $$KNOCK\ TEST$$
\begin{algorithmic} [1]

\STATE $\Rightarrow$ Identify target Node\\
\STATE $\Rightarrow$ Generate fake data packet with route via target node\\
\STATE $\Rightarrow$ Send packet to target node and wait for its \textsc{pack}\\
\IF {(\textsc{pack} is found)}
\STATE $\Rightarrow$ test Passed\\
\STATE $\Rightarrow$ Set reputation to default\\

\ELSE 
\STATE $\Rightarrow$ test Failed\\
\STATE $\Rightarrow$ Declare node as malicious and broadcast Warning message\\
\ENDIF

\end{algorithmic}

\indent Knock test is designed specifically for immediate neighbors to test whether a particular node is malicious or not and is only performed on nodes in suspicious state. In this test a dummy data packet with time to live (\textsc{TTL}) equal to 2 is sent to a node in question via last known route through that node.

The sender node overhears traffic of the node in question in promiscuous mode. If the node on which knock test is being performed successfully forwards the test packet to next hop then its reputation is set to default. In case it fails, then it is immediately put into malicious category and a warning message is broadcasted in the immediate neighborhood. If in case, the dummy packet is genuinely dropped because of bad channel conditions the node may be classified as malicious. However, it still has an opportunity to become a part of network again through redemption and fading mechanism, as explained earlier. This is done because the system only trusts first hand information for putting a node into malicious category, thus, giving self observation the highest weightage. The weightage assigned to warning message and avoid list citation is comparatively less than self observation.

\section{Implementation/Simulation}
This section first describes the simulation environment and then we compare the throughput of our system in the presence of malicious nodes against a defenseless \textsc{dsr} protocol. The network simulator ns2 (version 2.29) \cite{rNS} was used to run the simulations.
 Mobility of nodes is characterized by a mobility model, speed and `pause time'. The random waypoint model is selected as a mobility model in a 1000 $\times$ 1000 m$^2$ rectangular field. Using this mobility model, each movement is a straight line between a start and an arrival point, covered at a constant speed 
which is a uniform distribution, between 0 and 10 m.s$^{-1}$ for each movement. 
The pause time is the time period between two consecutive movements. Thus, the higher the pause time, lesser is the node's mobility. We have used 5 different pause times: 0, 100, 300, 600 and 900 seconds.\\
\indent There are two setups having a total of 10 and 20 nodes, with number of malicious nodes between 10 to 100\%. We use maximum 5 and 10 \textsc{cbr} (Constant Bit Rate) connections for 10 and 20 nodes respectively, sending 64 bytes packets with a 4 pkts.s$^{-1}$ sending rate. The bandwidth is 2 Mb.s$^{-1}$. The Medium Access Control (\textsc{Mac}) protocol used is \textsc{Ieee} 802.11.\\
 The malicious nodes are of the following nature: dropping an average of 99\% of the \textsc{cbr}-connection packets (data packets). The dropping decision is taken depending upon a number generated at random.  
We assume that malicious nodes do not drop the \textsc{dsr} routing packets like route request, route reply or error as they always want to be part of network. A malicious node dropping all the packets is comparatively less dangerous for the \textsc{manet} because in that case, it would drop all packets including routing packets. Following which, they would never be able to include themselves in any the communication routes. The fixed parameters for the simulation are listed in Table 1.
\begin{center}
\textbf{Table 1: Fixed Parameters}\\[5pt]
\begin{tabular}{|l|l|}
	\hline
	Parameter & Level \\
	\hline
		Area & 1000 m $\times$ 1000 m \\
		Speed & uniformly distributed between\\ & 0 and 10 m.s$^{-1}$ \\
		Radio Range & 250 m \\
		Placement & uniform \\
		Movement & random waypoint model \\
		MAC & 802.11 \\
		Sending capacity & 2 Mbps \\
		Application & CBR \\
		Packet size & 64 B \\
		Simulation time & 900 s \\
		\hline
	\end{tabular}
\end{center}
Thus, if no route has been established containing these nodes, they would never be able to drop any data packet sent to them either. As a consequence, they would not affect the throughput of whole network.
 For evaluating the performance of our system, we account for the \emph{Packet Delivery Ratio} and \emph{Routing Overhead} metrics. Packet delivery ratio is calculated as the ratio of data packets received to data packets sent. For routing overhead we have taken a ratio of number of control packets generated (request, reply and error) to the number of data packets sent thus, being basically a cost v/s. gain ratio. The routing overhead ratio gives us the approximate number of control packets for each data packet sent which should not be significantly greater as compared with that of normal \textsc{dsr}. Throughput refers to the actual measured performance of the system when the delay is considered. In the simulation results, the metrics of throughput are related to the average value per node. Finally, the average delay shows the average one-way latency observed between transmitting and receiving a packet. \\
Unless otherwise specified, the experiments are repeated ten times in all cases with varying random seed. The seed influences the placement and movement of the nodes. The radio range, sending capacity, and \textsc{mac} have been chosen to represent a typical adhoc mobile node; the speed is uniformly distributed between 0 and 10 m/s to represent speed of user in fixed location, walking or running. The simulation time is chosen to be long enough to potentially roam the whole area and is set to 900 seconds.
The system was deployed and tested with following values of constant parameters
\begin{center}
\textbf{Table 2: Values of Constant Parameters}\\[5pt]
\begin{tabular}{|l|l|}
	\hline
	Constant & Value \\
	\hline
		Neutral Rating & 0 \\
		Suspicious Threshold & -35 \\
		Malicious Threshold & -50 \\
		Window Size & 1 second (Default) \\
		Self Observation Weightage &  -5\\
		Warning Message Weightage &  -2\\
		Avoid List Appearance Weightage &  -1\\
		Inactivity Timeout Period & 20 seconds \\
		\hline
	\end{tabular}
\end{center}
Finally, \textsc{cbr} has been chosen for generating the traffic. The scenario and traffic connections have been randomly generated using the cbrgen and setdest utility from \textsc{cmu}'s Monarch project \cite{CMUS}.

\section{Results and Performance Evaluation}

\begin{figure}[h]
\centering
\includegraphics[scale=0.4]{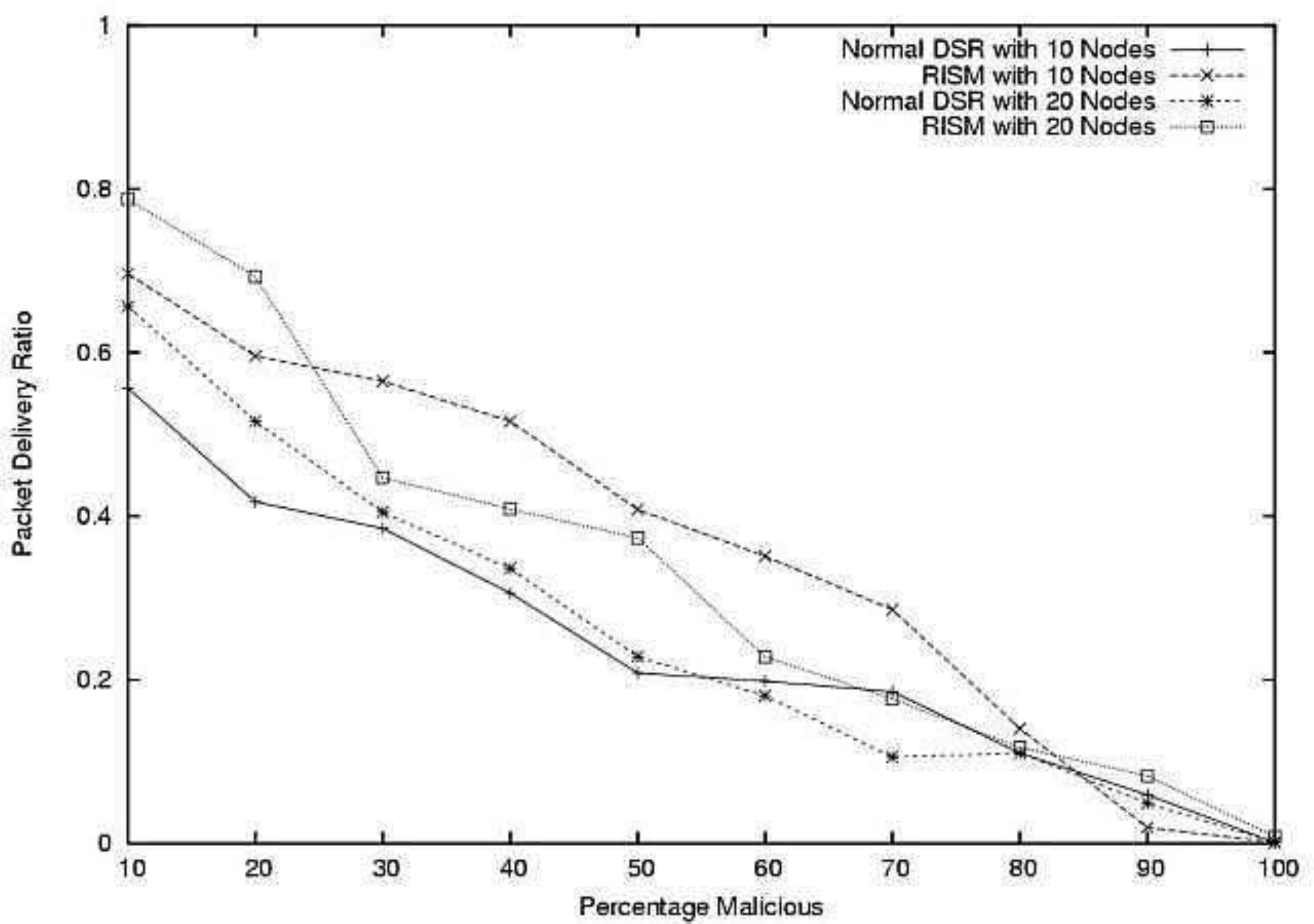}
\caption{\label{fig:PDratio}Packet Delivery Ratio Comparison}
\end{figure}
 Results for Packet Delivery Ratio are shown in the Figure~\ref{fig:PDratio}.
The system performs better than normal \textsc{dsr} comparing results taken after average over 10 iterations with different pause times. The system performance is significantly better than normal \textsc{dsr} when percentage of malicious nodes is less than 40. After which, it starts to deteriorate and significantly falls after 70\%. But, in case 70\% or more nodes are malicious we can simply discard the network as it is no longer of significance. There is no need to establish trust relationship and links among the nodes when 7 out of 10 are known to be malicious.

\begin{figure}[h]
\centering
\includegraphics[scale=0.4]{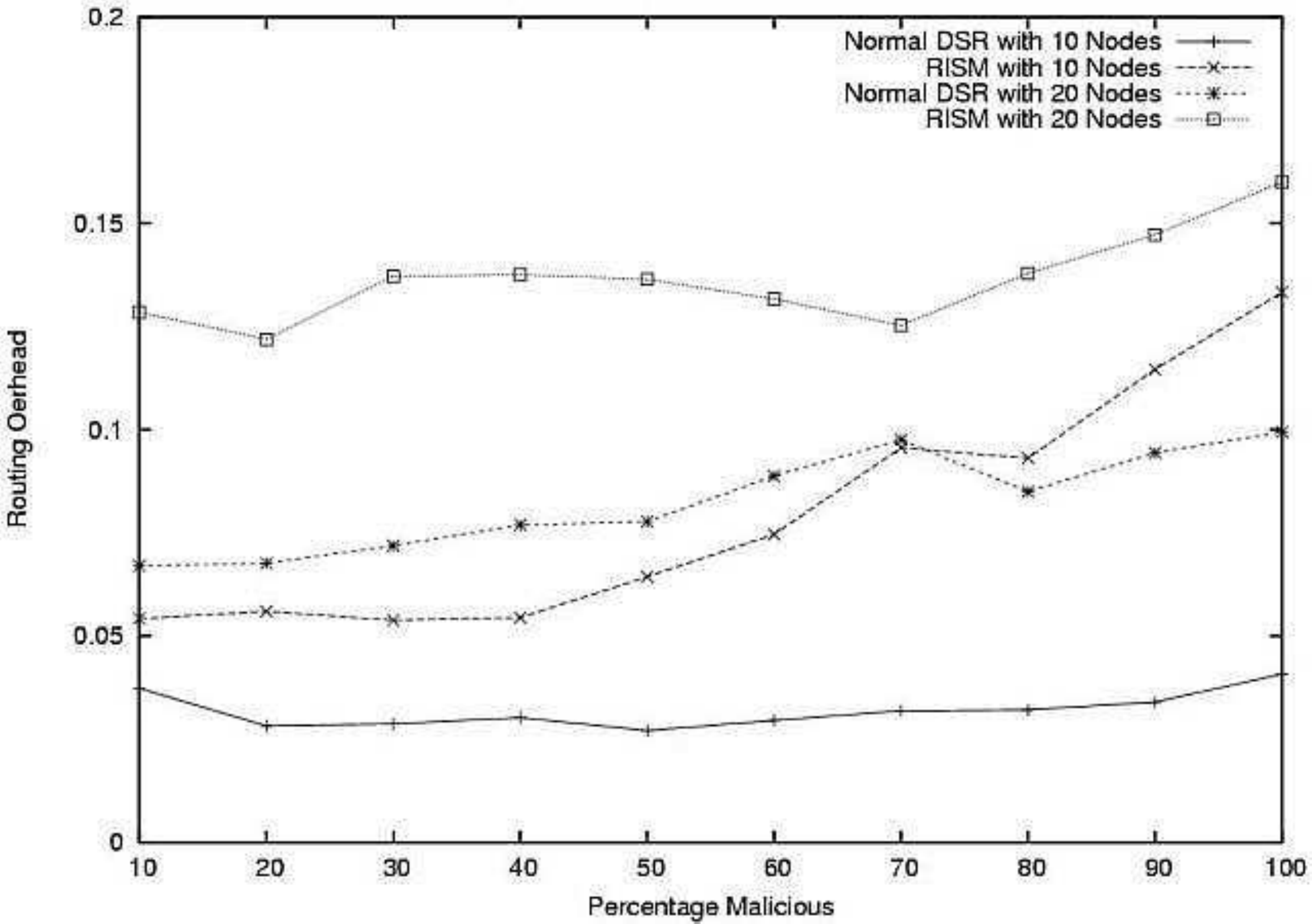}
\caption{\label{fig:ROcomp}Routing Overhead Comparison}
\end{figure}

\indent Figure~\ref{fig:ROcomp} shows routing overhead of our system protocol compared with normal \textsc{dsr}. Number of control messages in the network are significant, as more are the number of packets, more is the time spent in establishing routes and lesser is the number of data packets sent. Our system performs better than normal \textsc{dsr} without much extra routing overhead. This extra routing overhead is generated because whenever a new node is declared as malicious, a route error is generated and the link is broadcast as broken. After which, some more time is consumed to establish a new link. This is crucial to the \textsc{ids} performance.

\begin{figure}[h]
\centering
\includegraphics[scale=0.25]{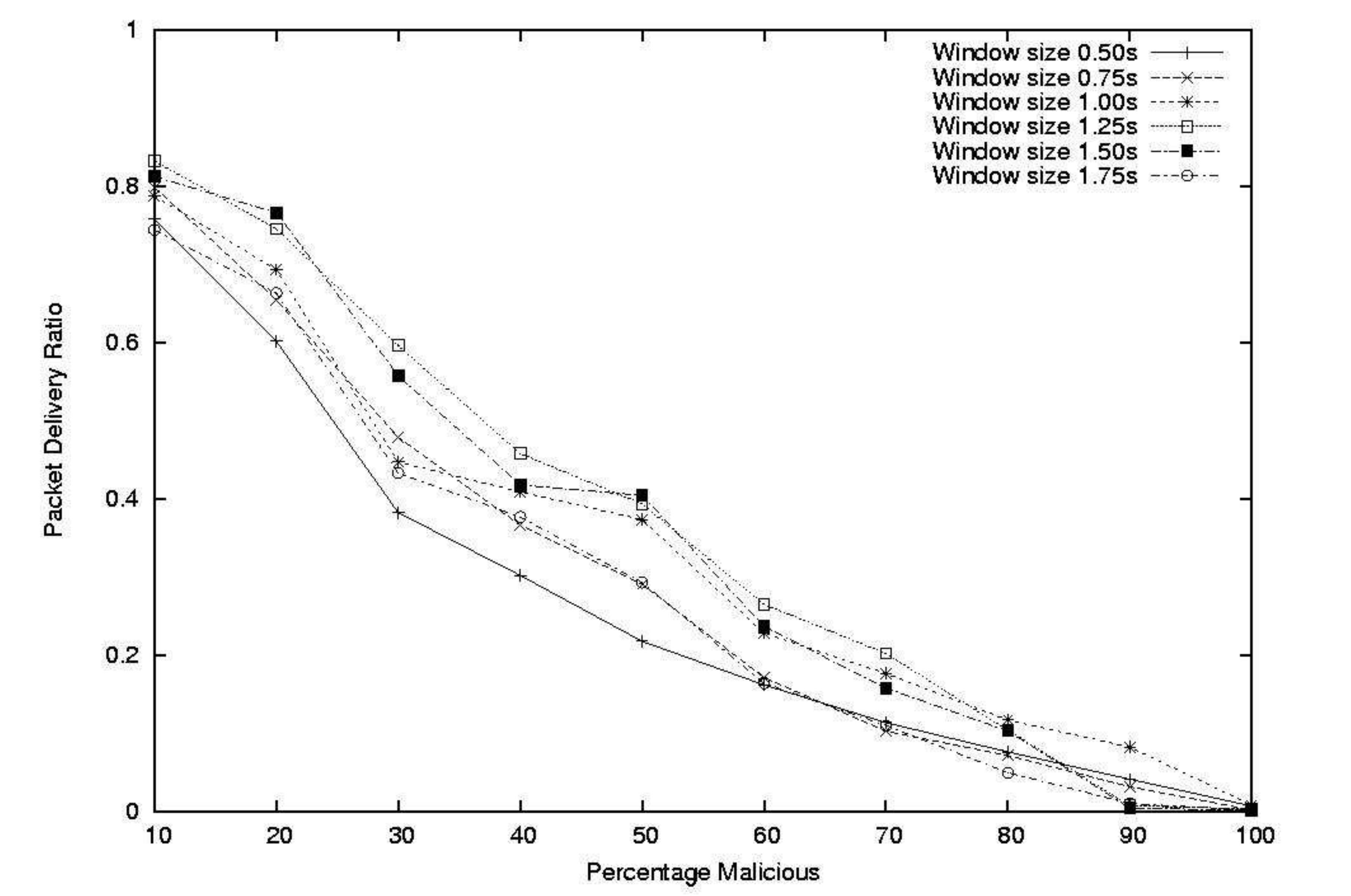}
\caption{\label{fig:Wsizes}System performance for various window sizes}
\end{figure}

\indent The Figure~\ref{fig:Wsizes} illustrates fine tuning of the system with respect to sliding window size. Window size is a crucial parameter for the system because of its role in deciding the system performance. As depicted by the figure, in most cases, the window size 1.25 seconds scenario delivers optimal packet delivery ratio as compared to other scenarios where window size is of 0.50, 0.75, 1.00, 1.50 and 1.75 seconds. From the figure, one can infer that for a small window size, the system is too busy in various book-keeping tasks for monitoring and reputation updating. For a larger window size, the system response gets too slow. Hence, the time to identify malicious nodes increases and accordingly does the number of packet drops. Therefore, overall performance of system deteriorates in terms of packet delivery ratio.  

\begin{figure}[h]
\centering
\includegraphics[scale=0.25]{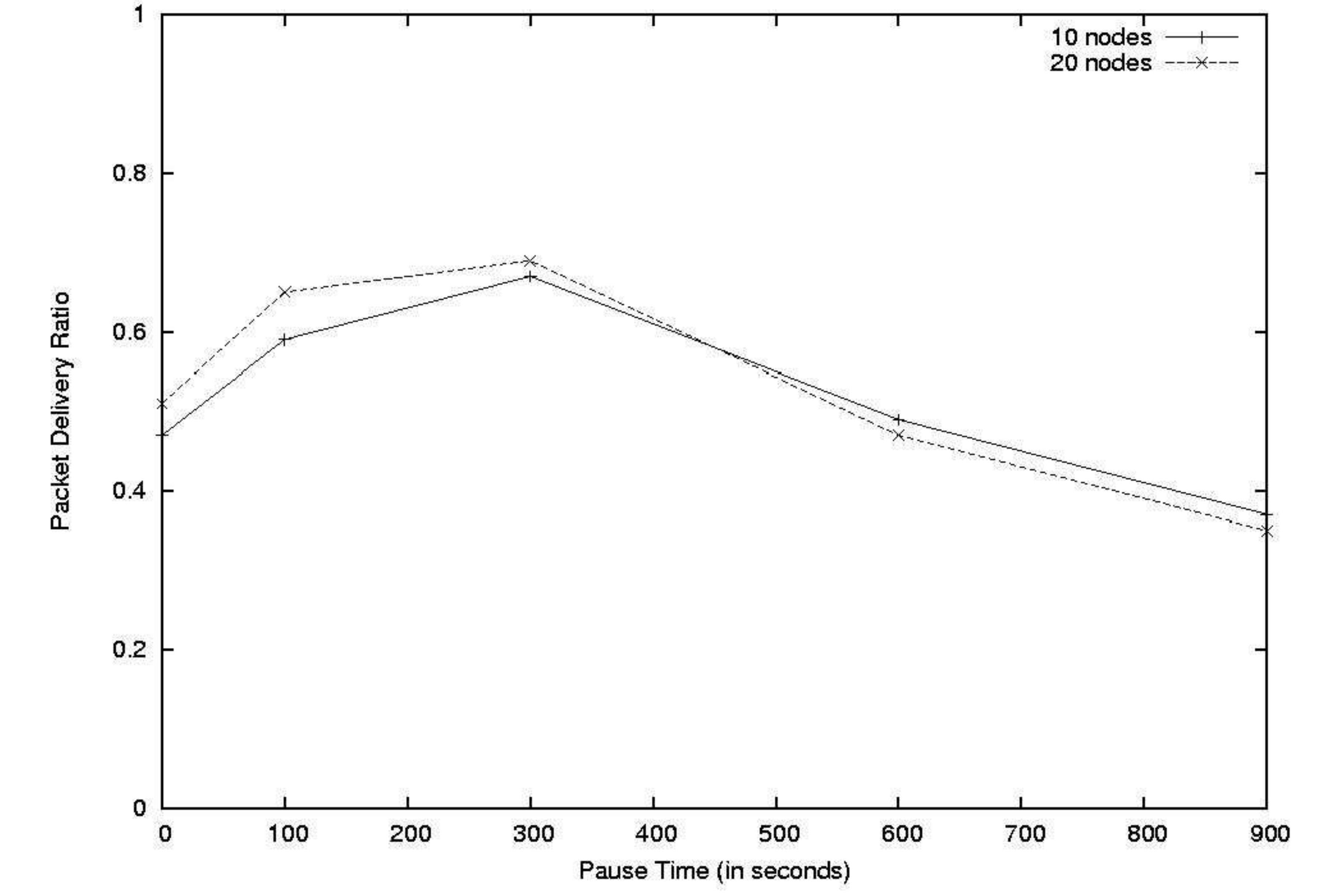}
\caption{\label{fig:Ptimes}System performance against Pause times}
\end{figure}

\indent The Figure~\ref{fig:Ptimes} shows packet delivery ratio values for the system against pause times of 0, 100, 300, 600 and 900 seconds.

 In a highly mobile scenario such as one with pause time of 0 seconds the system performance decreases as most of the time the system has to cope up with the mobility of nodes and tasks like updating route cache, discovering \& establishing routes etc. Likewise, in a static network scenario with pause times of 600 and 900 seconds, where the system does not have many choices in terms of clean routes, once these nodes get identified, the system performance also degrades.    

\begin{figure}[h]
\centering
\includegraphics[scale=0.3]{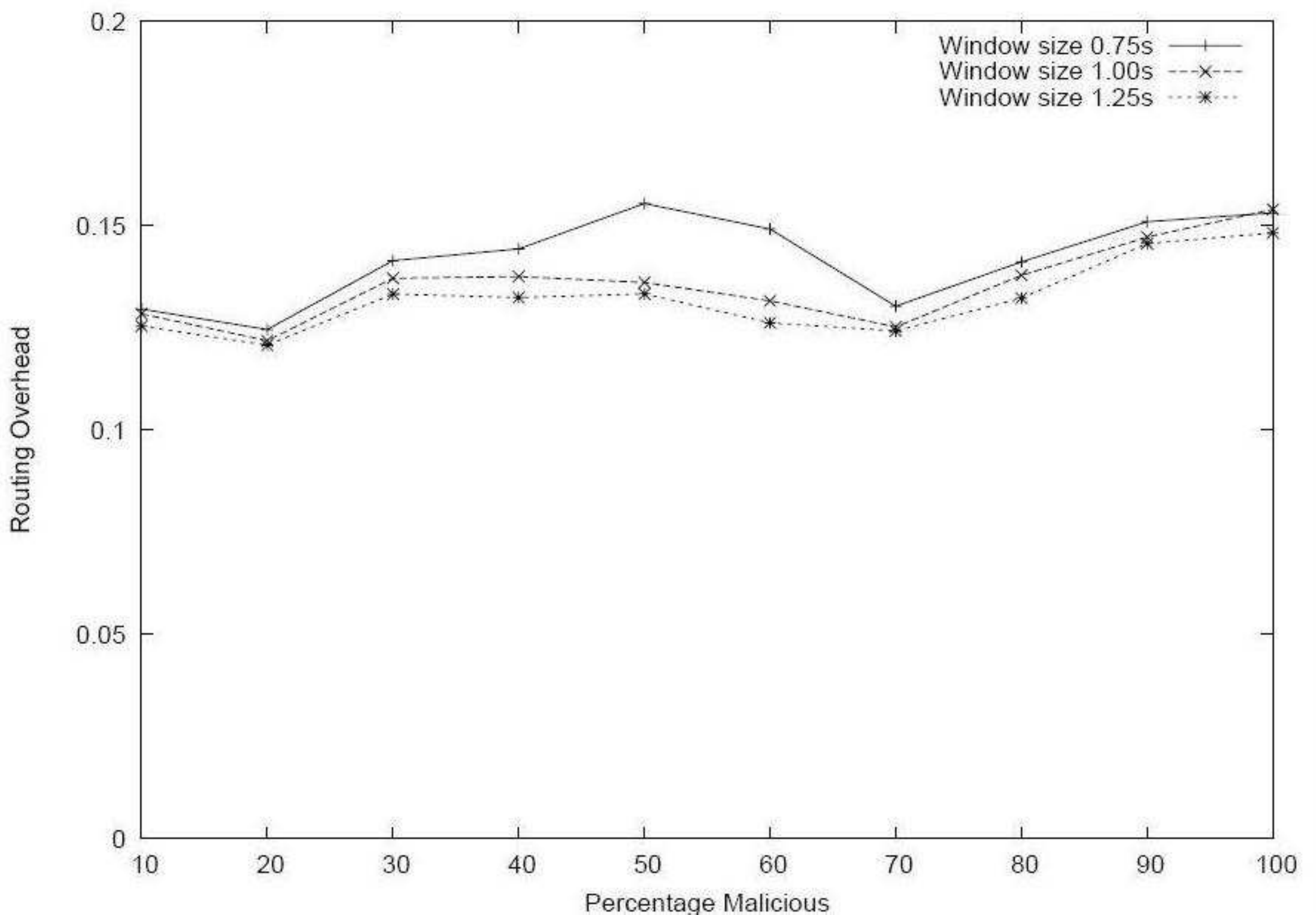}
\caption{\label{fig:ROWsizes}System routing overhead for various window sizes}
\end{figure}

As depicted by the Figure~\ref{fig:ROWsizes}, the performance is optimal for a scenario with window size of 1.25 seconds in terms of routing overhead. Although, the difference between various scenarios presented is not very significant, but it is crucial for system performance to optimize this value. 

In addition, the system also performs optimally in terms of the routing overhead incurred. Thus, the proposed solution is able to strengthen the defense of \textsc{dsr} protocol without incuring much of overhead.
 
\section{Conclusion And Future Perspective}
Mobile adhoc networks have a number of significant security issues which cannot be solved alone by simple \textsc{ids}. 
In this paper, we have critically examined the existing systems and outlined their strength and shortcomings. 
We have opted a semi-distributed approach for our system in terms of mode of information propagation among nodes. The goal was to design a system incorporating the best traits of all existing systems without incurring extra routing overhead. Congestion parameter, Knock test and Timing window are some new concepts that have been introduced in this system. Detailed simulations carried out over our system using ns2 for performance evaluation have contributed significantly to some crucial design issues. Optimal values of the parameters used are obtained and critically examined for efficient performance of the system. However, some additional study is required for evaluating the adequacy and importance of congestion parameter. The system performance can also be judged by interchanging the values of weightage assigned to self observation with that of other reputation update modes such as warning message and avoid list citations. It is our belief that some interesting results are bound to come with such studies which shall justify the system design in its current stage.
\balance

\break
\textbf{Animesh Kumar Trivedi} is a senior undergraduate student
of Information Technology at the Indian Institute of Information
Technology, Allahabad, India. His research interests include
Security issues in Wireless Computer Networks, Computer
Architecture and Distributed Systems. Further details about him can be had from \href{http://animesh.trivedi.googlepages.com/}{http://animesh.trivedi.googlepages.com/}\\

\textbf{Rajan Arora} is a senior undergraduate student of
Information Technology at the Indian Institute of Information
Technology, Allahabad, India. His research interests include
Wireless Computer networks, Cryptography, Network Security
and Distributed systems. Further details about him can be had from \href{http://arorarajan.googlepages.com/}{http://arorarajan.googlepages.com/}\\

\textbf{Rishi Kapoor} is a senior undergraduate student at the Indian Institute of Information
Technology, Allahabad, India pursuing his undergraduation
in Information Technology. His research interests include
Computer and Wireless networks, Network security and 
Databases. Further details about him can be had from \href{http://profile.iiita.ac.in/rkapoor_b03/}{http://profile.iiita.ac.in/rkapoor\_b03/}\\

\textbf{Sudip Sanyal} is an Associate Professor at the Indian Institute of Information Technology, Allahabad, India. He recieved his M.S. and Ph.D. degree from the Banaras Hindu University at Varanasi, India. His current activities lie in fields of Computer networks, Software Engineering, Parallel Computing and Natural language Processing. He has published numerous papers in various national and international journals and attended many conferences.\\

\textbf{Sugata Sanyal} is in the Faculty of the Tata Institute of Fundamental Research,
India. He received his Ph.D. degree from Mumbai University, India, M.Tech from IIT,
Kharagpur, India and B.E. from Jadavpur University, India. His current research
interests include security in wireless and mobile ad hoc networks, distributed
processing, and scheduling techniques. He has published numerous papers in national
and international journals and attended many conferences. He is in the editorial
board of three International Journals. He is co-recipient of ''Vividhlaxi Audyogik
Samsodhan Vikas Kendra Award (VASVIK)'' for Electrical and Electronics Science and
Technologies (combined) for the year 1985. He was a Visiting Professor in the
Department of Electrical and Computer Engineering and Computer Science in the
University of Cincinnati, Ohio, USA in 2003. He delivered a series of lectures and
also interacted with the Research Scholars in the area of Network Security in USA,
in University of Cincinnati, University of Iowa, Iowa State University and Oklahoma
State University. He has been an Honorary Member
of Technical Board in UTI (Unit Trust of India) and SIDBI (Small Industries
Development Bank of India). He has also acted as a consultant to a number of leading
industrial houses in India. More information about his activities is available at
\href{http://www.tifr.res.in/\~sanyal}{http://www.tifr.res.in/\~sanyal}.


\begin{thebibliography}{0}
\bibitem{rDSDV} C. E. Perkins, P. Bhagwat: ``Highly Dynamic Destination-Sequenced Distance Vector (DSDV) for Mobile Computers" \textit{Proc. of the SIGCOMM 1994 Conference on Communications Architectures, Protocols and Applications}, pp 234-244, Aug (1994).
\bibitem{rWRP} S. Murthy, J.J. Garcia-Luna-Aveces: ``A Routing Protocol for Packet Radio Networks", \textit{Proc. ACM International Conference on Mobile Computing and Networking}, pp. 86-95, November, (1995). 
\bibitem{rCGSR} Ching-Chuan Chiang, Hsiao-Kuang Wu, Winston Liy, Mario Gerla: ``Routing in Clustered Multihop, Mobile Wireless Networks with Fading Channel", \textit{IEEE Singapore International Conference on Networks, (SICON'97)}, pp. 197-211, Singapore, 16.-17. April (1997).
\bibitem{rDSR}David B. Johnson, David A. Maltz, and Josh Broch, ``DSR The Dynamic Source Routing Protocol for Multihop Wireless Ad Hoc Networks" In \textit{Ad Hoc Networking}, edited by Charles E. Perkins, chapter 5, pages 139--172. Addison-Wesley, (2001).
\bibitem{rAODV}Charles E. Perkins and Elizabeth M. Royer: ``Ad-hoc On-Demand Distance Vector Routing" in \textit{Proceedings of the 2nd IEEE Workshop on Mobile Computing Systems and Applications}, New Orleans, LA, pp. 90-100, February (1999).
\bibitem{rTORA} V. Park, S. Corson: ``Temporally-Ordered Routing Algorithm (TORA)" VERSION 1 Internet Draft, draft-ietf-manet-tora-spec- 03.txt, June (2001).
\bibitem{rABR} Chai-Keong Toh: ``A Novel Distributed Routing Protocol To Support Ad hoc Mobile Computing", \textit{Proc. IEEE 15th Annual International Phoenix Conference on Computers and Communications}, IEEE IPCCC 1996, 27 March-29, Phoenix, AZ, USA, pp. 480-486 (1996).
\bibitem{rSSR} R. Dube, C. D. Rais, K. Wang and S. K. Tripathi: ``Signal Stability based adaptive routing (SSR alt SSA) for ad hoc mobile networks", \textit{IEEE Personal Communication}, Feb. (1997).
\bibitem{rZRP} Zygmunt J. Haas, Marc R. Pearlman, Prince Samar: ``The Zone Routing Protocol (ZRP) for Ad Hoc Networks", Internet Draft, http://www.ietf.org/proceedings/02nov/I-D/draft-ietf-manet-zone-zrp-04.txt, work in progress, July (2002).
\bibitem{WDOG}Sonja Buchegger, Cedric Tissieres, Jean-Yves Le Boudec, ``A Test-Bed for Misbehavior Detection in Mobile Ad-hoc Networks — How Much Can Watchdogs Really Do?,"  \textit{Sixth IEEE Workshop on Mobile Computing Systems and Applications (WMCSA'04)}, pp. 102-111, (2004).
\bibitem{rCORE}P. Michiardi, R. Molva, ``Core: A COllaborative REputation mechanism to enforce node cooperation in Mobile Ad Hoc Networks", \textit{Institut EurecomResearch Report} RR-02-062 - December (2001).
\bibitem{rCONF}Sonja Buchegger and Jean-Yves Le Boudec, ``Performance Analysis of the \textsc{confidant} Protocol: Cooperation Of Nodes:Fairness In Dynamic Ad-hoc NeTworks" \textit{Proceedings of IEEE/ACM Symposium on Mobile Ad Hoc Networking and Computing (MobiHOC)}, Lausanne, June(2002).
\bibitem{rOCEAN}Sorav Bansal and Mary Baker, ``Observation based cooperation enforcement in ad hoc networks" Technical Report, Stanford University, arXiv:cs.NI/0307012 v2 6 Jul (2003).
\bibitem{rTRUST}P. Resnick and R. Zeckhauser. ``Trust among strangers in internet transactions: Empirical analysis of ebay's reputation system" In M. R. Baye, editor,  \textit{The Economics of the Internet and E-Commerce}, volume 11 of Advances in Applied Microeconomics. Amsterdam, Elsevier Science, (2002).
\bibitem{FACC} Sonja Buchegger, Jean-Yves Le Boudec: ``Coping with False Accusations in Misbehavior Reputation Systems for Mobile Ad-hoc Networks" EPFL Technical Report IC/2003/31 (2003).
\bibitem{SHAND}S. Buchegger and J.-Y. Le Boudec, ``The effect of rumor spreading in reputation systems for mobile ad-hoc networks" \textit{Proc. WiOpt'03(Modeling and Optimization in Mobile Ad Hoc and Wireless Networks)}, (2003).
\bibitem{rRISM} Animesh K. Trivedi, Rishi Kapoor, Rajan Arora, Sudip Sanyal and Sugata Snayal: ``RISM - Reputation Based Intrusion Detection System for Mobile Adhoc Networks", accpeted in CODEC'06, Kolkata, India (http://www.irpel.org/phpfiles/codec-06.php) to be held in Dec. (2006).
\bibitem{NTRAL}P. Yau and C. J. Mitchell, ``Reputation methods for routing security for mobile ad hoc networks", in \textit{Proceedings of SympoTIC '03}, Joint IST Workshop on Mobile Future and Symposium on Trends in Communications, Bratislava, Slovakia, October (2003).
\bibitem{rNS} NS Home Page: http://www.isi.edu/nsnam/ns/
\bibitem{CMUS}``CMU Monarch Project web site." http://www.monarch.cs.cmu.edu/
\bibitem{REPN}Sonja Buchegger and Jean-Yves Le Boudec, ``Self-Policing Mobile Ad-Hoc Networks by Reputation Systems" \textit{IEEE Communication Magazine}, vol. 43, num. 7, p. 101(2005).
\bibitem{WOG}Sergio Marti, T.J. Giuli, Kevin Lai, and Mary Baker, ``Mitigating routing misbehavior in mobile ad hoc networks" \textit{Proceedings of the 6th annual international conference on Mobile computing and networking} Boston, Massachusetts. Pages: 255 - 265 (2000). 
\end{thebibliography}
\end{document}